# The method of exchange perturbation theory as applied to magnetic ordering in high-*Tc* materials

Elena V. Orlenko* and Tatiana Latychevskaia
St. Petersburg State Technical University,
Polytechnicheskaya St. 29, 195251 St. Petersburg, Russia
*eorlenko@mail.ru

## Abstract
We unify the method of exchange perturbation theory for multicenter systems. For the case of exchange degeneracy in the total spin of the system we give a secular equation that is more compact and convenient for calculations than those obtained earlier. On the basis of this formalism we develop an algorithm for calculating the Heisenberg parameter for magnetic materials. Finally, we calculate the characteristics of antiferromagnetic transitions for the high-*Tc* materials $La_{2-x}MeCuO_4$ and $YBa_2Cu_3O_6$.

## Introduction

Usually the microscopic description of magnetic materials is either a statistical analysis of spin systems on the basis of the Heisenberg equation or a calculation and analysis of magnetization and magnetic susceptibility in the single electron approximations in models of the Stoner [1] or Hubbard [2] type. But, in one way or another, the main parameter characterizing the spin system is still the exchange integral, which is chosen differently in different models; for instance, in Hubbard-type models it is the Coulomb exchange interaction of electrons strongly localized at the centers and, therefore, calculated in the Wannier-function representation. The Heisenberg parameter in spin models is estimated semiphenomenologically, by reduction to the simplest Heitler–London two-center problem [3]. Clearly, the intercenter interaction of electrons belonging to the inner shells of ions, which is responsible for the spontaneous orientation of the spins, is much more complicated than in the models. First, the overlap of the atomic wave functions of the inner electrons belonging to different centers is responsible for their spin correlation; thus, the use of a ''truncated'' Wannierfunction basis artificially reduces the contribution of the intercenter exchange to the interaction energy and in this way essentially eliminates the intercenter correlation effects. Second, for many materials the number of ''active'' electrons of the inner shells of atoms participating in intercenter interaction exceeds unity, so that the wave function describing at least a two-center system is more than two-particle, and its spatial part is not reduced only to symmetric or antisymmetric form, as it is in the Heitler–London model. The use of the Slater determinant, which incorporates both coordinate and spin one-electron states simultaneously, makes it impossible to analyze the spin state emerging as a result of the interaction. Third, often the magnetic orientation of the spins is caused not simply by two-center exchange but by superexchange, in which the electrons of three or more centers participate [4, 5]. The constants of such interaction are approximated variationally by combinations of pair integrals,6 which actually means that the nonadditive part specific to multicenter interaction is discarded.

The situation is such that a meaningful description of spin systems requires not only effective summing over the states of the possible spin configurations (an enormous number of fine papers have been written on the subject, including those that use the renormalization method) but also developing an algorithm that would allow doing consistent calculations of the fundamental parameter present in any statistical scheme, the Heisenberg parameter. The discovery of anomalous magnetic effects in high-$Tc$ superconductors is vivid proof of the necessity of developing such an algorithm, since these effects are caused not so much by structural transformations in the crystal as by the change in the nature of the exchange interaction proper. The point is that such crystals as $YBa_2Cu_3Cu_3O_6$ and $La_2CuO_4$ and the materials $Rb_2MnF_6$ and $Rb_2CoF_4$ isostructural to the latter, in the pure or stoichiometric state are antiferromagnetic insulators with a fairly high transition temperature. Alloying, which is done by replacing the $La^{2+}$ ion by an atom of a metal with valence 2+ (such as $Cu^{2+}$, $Ba^{2+}$, and $Sr^{2+}$, so that we have the alloys $La_{2-x}Me_xCuO_4$, $Rb_2Mn_{1-x}Mg_xF_4$, and $Rb_2Co_{1-x}Mg_xF_4$, respectively) or by changing the oxygen content ($La_2CuO_{4-\delta}$ and $YBa_2Cu_3O_6$), lowers the Néel temperature so drastically that the antiferromagnetic state may be destroyed [7–9] and replaced by a weakly fluctuating 3D state of a spin liquid with preferentially parallel pair orientation of the spins.

A theoretical analysis of the behavior of these systems yields contradictory results. For example, according to calculations done with the one-electron band approximation, the $La_2CuO_4$ compound is a nonmagnetic metal.10,11 At the same time, electronic-structure models used in studies of the Heisenberg Hamiltonian for two- and three-center systems in the representation of the spin eigenfunctions of the operator $J(\hat{s}_1 + \hat{s}_2)\hat{s}_3$ provide a fairly realistic phase diagram that describes the transition of the system from the 2D antiferromagnetic state into the 3D state of a spin liquid and then into the superconducting state [12]. However, in these papers the numerical value of J for different spin configurations is estimated by analyzing experimental data.

The Heisenberg parameter can in principle be calculated consistently by applying the formulas of exchange perturbation theory, which takes into account the effects of intercenter overlap of the wave function. There are many formal variants of this theory, which are classified according to the way in which the algorithm is constructed (a detailed description of this classification can be found in Kaplan's monograph [13], which also analyzes the merits and drawback of the variants). There are two problems that must be dealt with in constructing the algorithm of exchange perturbation theory: the nonorthogonality of the base of the multicentersystem wave functions, which are antisymmetric in intercenter permutations (this problem is related to what is known as the overfilling catastrophe), and the asymmetry of the perturbation operator and the unperturbed part of the Hamiltonian with respect to intercenter permutations of electrons. In other words, if $[H, A] = 0,$ where $H = H^0 + V$ is the total system Hamiltonian, $\hat{H}_0$ is its unperturbed part, and A is the antisymmetrization operator, then $[H^0, A] \neq 0,$ and $[V, A] \neq 0$. The $H^0$ zeroth wave function antisymmetrized in intercenter permutations is not an eigenfunction of $H^0$, and the corrections calculated in the perturbation V contain nonphysical contributions.

All variants of exchange perturbation theory can be divided into two groups [13]. The first consists of theories nonsymmetric in the Hamiltonian. The second consists of approaches that make it possible to use the common Rayleigh-Schrödinger perturbation theory by setting up a special zeroth symmetric Hamiltonian for which the antisymmetric functions are eigenfunctions. The first group uses the fundamental basis of zeroth functions nonsymmetric in intercenter permutations, functions that are the eigenfunctions of the nonsymmetric Hamiltonian $H^0$. Antisymmetrization is done post factum at each interpolation step, which in the final analysis requires using a variational procedure in the perturbation theory formalism (as, say, is done in Ref. [14]). An attempt to modify the Hamiltonian so that the perturbation operator becomes symmetric (this is known as the Sternheimer procedure [15, 16]) leads to a non-Hermitian total Hamiltonian and actually limits the use of the method to two-electron systems.

The second group of variants of exchange perturbation theory can be assumed to include the work of Ritchie [17]. In this paper special projection operators are employed whose action on an antisymmetric function is reduced to selecting a term with a specific permutation. Since the explicit form of these operators was not given, it was assumed that they make the Hamiltonian non-Hermitian. Rumyantsev [18] demonstrated the effectiveness of using such symmetrization of the Hamiltonian. Despite a conceptual difficulty, a variant of exchange perturbation theory the Rayleigh-Schrödinger form was constructed, and on the basis of this theory the spectral characteristics of the hydrogen–helium system were calculated with high accuracy. Only in Ref. [19] were the projection operators derived explicitly. These operators symmetrize the perturbation operator and the unperturbed Hamiltonian and retain the hermiticity of the total Hamiltonian in the sense that its eigenvalues are real numbers. In the same paper the variant of exchange perturbation theory was generalized to the case where the system is degenerate in total spin. A detailed description of the variant and a broad range of applications used in calculating specific system, including spin system, can be found in Ref. [5]. Unfortunately, the organization of the material in Refs. [5] and [19] makes it impossible to explicitly analyze all the small parameters of the theory in which the power series expansions are done, and this makes the use of the formulas difficult. We also note that in Ref. [19] the corrections to the energy when degeneracy in total spin is lifted were calculated only for the case where there is another degeneracy, in orbital momentum.

In the present paper we use the idea of the method of symmetrizing the Hamiltonian [17–19] and construct a more compact algorithm of exchange perturbation theory. This allowed us to estimate the smallness of the terms discarded at each iteration step, terms that emerge because of the overfilling of the nonorthogonal base of antisymmetric functions. Due to a change in the normalization condition for the antisymmetric functions, all the projections in this variant, including Ritchie-type operators, are simpler. This has made it possible to obtain a solution of the secular equation when the system is degenerate in total spin in a more general form, more suitable for calculations. To show the possibilities of our version of exchange perturbation theory with degeneracy, we examined the high-$Tc$ materials $La_{2-x}Sr_xCuO_4$ and $Ba_2Cu_3O_{6+x}$, for which we calculated the Heisenberg parameter in the stoichiometric and alloyed states. We show that alloying these materials dramatically changes the magnitude and sign of the exchange and superexchange integrals. These changes lead, in accordance with the models of Ref. [12], to destruction of the 2D antiferromagnetic state and emergence of a 3D ferromagnetic state of spin liquid.

## 1. Exchange perturbation theory

A system of noninteracting atoms in the adiabatic approximation can be specified solely by the electron part of the wave function, which is simply the product of atomic wave functions. We write the spatial part as

$$\Phi(r_1,\ldots r_N) = \prod_\alpha \psi_\alpha(r_i,\ldots r_j), \tag{1.1}$$

where $\alpha$ is the number of a center, or atom, and $i,\ldots, j$ are the numbers of the electrons belonging to an atom.

The Hamiltonian describing such a system consists of the kinetic energy of all the electrons, the potential energy of the interaction of electrons and the ''parent'' center, and the interaction of the electrons belonging to one center with each other.

If the Hamiltonian does explicitly contain spin operators, then

$$H^0 \Phi_n(r_1,\ldots r_N) = E_n^0 \Phi_n(r_1,\ldots r_N),$$

where $\{E_n^0\}$ is the set of the eigenfunctions of the energy of the noninteracting system, and $\Phi_n(r_1,\ldots r_n)$ are the eigenfunctions corresponding to this spectrum.

The distances between the centers are such that atomic wave functions may strongly overlap. Then, in accordance with the Pauli exclusion principle, even for a noninteracting system the complete wave function must be antisymmetrized. So that the spin part can easily be separated later, leaving only the spatial part, Young tableaux can be used to antisymmetrize the complete wave function [20]. Then the spatial part of the complete wave function of the noninteracting system is

$$\Psi_n^0(r_1,\ldots r_N) = A\Phi_n(r_1,\ldots r_N), \tag{1.2}$$

where $A$ is the antisymmetrization operator corresponding only to the spatial part of the Young tableau or, in greater detail,

$$\Psi_n^0(r_1,\ldots r_N) = \frac{1}{f_n^P} \sum_{p=0}^{P} (-1)^{g_P} \Phi_n^p(r_1,\ldots r_N), \tag{1.3}$$

where $p$ is the number of permutations, $g_P$ is the parity of that permutation, $P$ is the total number of possible intercenter permutations, $1/f_n^P$ is the normalization constant, and $\Phi_n^p(r_1,\ldots r_N)$ is the wave function of the form (1.1) containing the permutation $p$ that corresponds to the Young tableau. We find the normalization constant (1.3) from the condition

$$\langle \Phi_n^0 | \Psi_n^0(r_1,\ldots r_N) \rangle = 1. \tag{1.4}$$

Then

$$f_n^P = \sum_{p=0}^{P} (-1)^{g_P} \left( \Phi_n^0 | \Phi_n^p \right) \tag{1.5}$$

differs by a factor $\sqrt{P}$ from the same constant in the normalization $\langle \Psi^0 | \Psi^0 \rangle = 1$ commonly used in exchange perturbation theory.

We introduce a projection operator $\Lambda_n^\pi$ that separates the term with the πth permutation in an antisymmetrized function of type (1.3):

$$\Lambda_n^\pi = | \Phi_n^\pi )( \Phi_n^\pi |. \tag{1.6}$$

Then

$$\Lambda_n^\pi | \Psi_n^0 \rangle = (-1)^{g_\pi} | \Phi_n^\pi ).$$

Now we can write the system Hamiltonian without a perturbation in a form invariant under intercenter permutations:

$$H_0 = \frac{1}{f_n^P} \sum_{p=0}^{P} H_p^0 \Lambda_n^p , \quad \widehat{V} = \frac{1}{f_n^P} \sum_{p=0}^{P} V_p \Lambda_n^p, \tag{1.7}$$

where $H_p^0$ and $V_p$ are the unperturbed Hamiltonian and the perturbation corresponding to the $p$th intercenter permutation of the electrons.

As usual, the perturbation operator incorporates the interaction of the nuclei, the interaction of electrons with ''foreign'' nuclei, and the interaction of the electrons belonging to different centers. Now the zeroth antisymmetric wave function (1.3) is the eigenfunction of the invariant unperturbed Hamiltonian $H_0$ (1.7):

$$H_0 \Psi_n^0 = E_n^0 \Psi_n^0. \tag{1.8}$$

The Hamiltonian of the interacting system is always invariant under electron perturbations, so that the solution of the Schrödinger equation can be antisymmetric under any electron permutation:

$$H\Psi = E\Psi. \tag{1.9}$$

Solving Eq. (1.9) by the method of successive approximations, we seek the perturbative corrections to the zeroth energy and wave function taken from (1.8). For instance, for the initial wave function and its corrections to have the proper symmetry, we use the perturbation operator and the unperturbed part of the Hamiltonian in the form of (1.7).
Then instead of (1.9) we have

$$(H_0 + V)\Psi_i = E_i \Psi_i, \tag{1.10}$$

where

$$\Psi_i = \Psi_i^{(0)} + \Psi_i^{(1)} + \ldots$$
$$E_i = E_i^{(0)} + E_i^{(1)} + \ldots.$$

At the beginning we keep only the zeroth- and first-order terms in (1.10). Then, allowing for (1.8), we have

$$H_0 \Psi_i^{(1)} + V \Psi_i^{(0)} = E_i^{(1)} \Psi_i^{(0)} + E_i^{(0)} \Psi_i^{(1)}. \tag{1.11}$$

We impose the intermediate normalization condition

$$\langle \Phi_i | \Psi_i \rangle = \langle \Phi_i | \Psi_i^0 \rangle \tag{1.12}$$

i.e., $\langle \Phi_i | \Psi_i - \Psi_i^0 \rangle = 0$. This means that all corrections to the wave function of the zeroth approximation lie in the subspace of the state vectors orthogonal to $|\Psi_i^0\rangle$.

Let us introduce the projector on the subspace of vectors parallel to $|\Psi_i^0\rangle$

$$P_i = |\Psi_i^0\rangle\langle\Phi_i|, \tag{1.13}$$

where $P_i|\Psi_i^0\rangle \equiv |\Psi_i^0\rangle$. Since

$$P_i H_0 |\Psi_i^{(1)}\rangle = \frac{|\Psi_i^0\rangle}{f_i^P} \sum_{p=0}^{P} \langle\Phi_i^0|H_0^p|\Phi_i^p\rangle\langle\Phi_i^p|\Psi_i^{(1)}\rangle =$$
$$E_i^0 \frac{|\Psi_i^0\rangle}{f_i^P} \sum_{p=0}^{P} \langle\Phi_i^0|\Phi_i^p\rangle\langle\Phi_i^p|\Psi_i^{(1)}\rangle = E_i^0 |\Psi_i^0\rangle\langle\Phi_i^0|\Psi_i^{(1)}\rangle = E_i^0 P_i |\Psi_i^{(1)}\rangle \tag{1.14}$$

after the operator (1.13) is applied to Eq. (1.11) we get $\langle\Phi_i|V|\Psi_i^0\rangle|\Psi_i^0\rangle = E_i^{(1)}|\Psi_i^0\rangle$. This leads to an expression for the first-order correction to the energy,

$$E_i^{(1)} = \langle\Phi_i|V|\Psi_i^0\rangle. \tag{1.15}$$

Now let us introduce the projector on the subspace of vectors orthogonal to $|\Psi_i^0\rangle$ in accordance with the property of the double vector product:

$$O_i = 1 - P_i \quad \text{where} \quad O_i|\Psi_i^0\rangle \equiv 0 \tag{1.16}$$

or, to put it differently,

$$O_i|\Psi\rangle = \langle\Phi_i|\times|\Psi\rangle \times |\Psi_i^0\rangle \tag{1.17}$$

i.e., $|\Psi\rangle = |\Psi_i^0\rangle\langle\Phi_i|\Psi\rangle + \langle\Phi_i|\times|\Psi\rangle \times |\Psi_i^0\rangle$.

From (1.13) and (1.17) we see that the antisymmetric basis of zeroth wave functions is actually only weakly nonorthogonal,

$$P_i|\Psi_n^0\rangle = |\Psi_i^0\rangle\langle\Phi_i^0|\Psi_n^0\rangle = |\Psi_i^0\rangle \frac{1}{f_P} \sum_{p=0}^{P} \langle\Phi_i^0|\Phi_n^p\rangle(-1)^{g_p} \approx 0, \tag{1.18}$$

because $\langle\Phi_i^0|\Phi_n^0\rangle = 0$ and $\langle\Phi_i^0|\Phi_n^p\rangle \approx 0$ since the overlap of the wave functions of the ground and excited belonging to different centers is insignificant (this resembles the situation in diffraction theory for wave optics). Accordingly,

$$\langle\Phi_i|\times|\Psi_n^0\rangle \times |\Psi_i^0\rangle \approx |\Psi_n^0\rangle. \tag{1.19}$$

We seek, in accordance with (1.12), the first-order correction to the antisymmetrized function of the zeroth approximation in the form of an expansion:

$$\Psi_i^{(1)} = \sum_n C_n \Psi_n^0, \tag{1.20}$$

where $n \neq i$. Inserting the expansion (1.20) in (1.11) and applying the operator (1.16) to the result, we get

$$O_i V |\Psi_i^0\rangle = \sum_n C_n \left(E_i^0 - E_i^n\right)|\Psi_n^0\rangle, \quad (1.21)$$

where we have allowed for (1.18) and (1.19).

Using the property of completeness of the orthogonal basis of the nonsymmetric zeroth functions,

$$\sum_n |\Phi_n^p\rangle\langle\Phi_n^p| = 1$$

we can write the left-hand side of Eq. (1.21) as follows:

$$\frac{1}{P}\sum_{p=0}^{P}\sum_n |\Phi_n^p\rangle\langle\Phi_n^p|O_i V|\Psi_i^0\rangle = \frac{1}{P}\sum_{p=0}^{P}\sum_n |\Phi_n^p\rangle(-1)^p\langle\Phi_n^p|O_i V|\Psi_i^0\rangle = $$
$$= \frac{1}{P}\sum_n f_n^P |\Psi_n^0\rangle\langle\Phi_n^0|O_i V|\Psi_i^0\rangle. \quad (1.22)$$

Since

$$\left(\Phi_i|O_i V|\Psi_i^0\right) = \left(\Phi_i|1 - |\Psi_i\rangle\right)\left(\Phi_i|V|\Psi_i^0\right) \equiv 0,$$

we must drop the term with $n=i$ from (1.22).

Finally, Eq. (1.21) becomes

$$\frac{1}{P}\sum_n f_n^P |\Psi_n^0\rangle\left(\Phi_n^0|O_i V|\Psi_i^0\right) = \sum_n C_n \left(E_i^0 - E_n^0\right)|\Psi_n^0\rangle, \quad (1.23)$$

from which we find the first-order correction to the wave function:

$$|\Psi_i^{(1)}\rangle = \frac{1}{P}\sum_n f_n^P \frac{\left(\Phi_n^0|O_i V|\Psi_i^0\right)}{\left(E_i^0 - E_n^0\right)}|\Psi_n^0\rangle. \quad (1.24)$$

Higher-order corrections can be found similarly.

## 2. The case of degeneracy

The zeroth wave function for a multicenter system can be antisymmetrized by various Young tableaux, which differ for different values of the total spin of the system. In other words, a multicenter system of noninteracting electrons is degenerate in total spin, with the degeneracy lifted by allowing for ordinary intercenter interaction. Thus,

$$\Psi_{n\alpha}^0 = A_\alpha \Phi_n^0, \quad H_\alpha^0 \Psi_{n\alpha}^0 = E_n^0 \Psi_{n\alpha}^0, \quad H_\alpha^0 = \sum_{p=0}^{P}\frac{1}{f_n^\alpha}H_p^0 \Lambda_n^p, \quad (2.1)$$

where $\{\Psi_{n\alpha}^0\}$ is the set of wave functions antisymmetrized by different Young tableaux $\alpha$ and corresponding to the same energy level $E_n^0$ of the system.

We seek the wave function of an interacting multicenter system of electrons in the form

$$\Psi_i = \sum_\beta C_\beta^0 \Psi_{\beta i}^0 + \varphi. \tag{2.2}$$

If we substitute (2.2) in the complete Schrödinger equation (1.9), we obtain

$$H \sum_\beta C_\beta^0 \Psi_\beta^0 + H\varphi = \left(E_i^0 + \varepsilon\right) \sum_\beta C_\beta^0 \Psi_\beta^0 + \left(E_i^0 + \varepsilon\right)\varphi.$$

Since the total Hamiltonian is invariant under all permutations, it can be taken outside the summation sign in accordance with the Young tableau:

$$\sum_\beta \left(H_\beta^0 + \hat{V}_\beta\right) C_\beta^0 \Psi_\beta^0 + H\varphi = \left(E_i^0 + \varepsilon\right) \sum_\beta C_\beta^0 \Psi_\beta^0 + \left(E_i^0 + \varepsilon\right)\varphi. \tag{2.3}$$

Using (2.1), we can shift all the terms containing $\varphi$ to the left-hand side of Eq. (2.3) and all the other terms to the right-hand side. The result is

$$H\varphi - \left(E_i^0 + \varepsilon\right)\varphi = \sum_\beta \left(\varepsilon - V_\beta\right) C_\beta^0 \Psi_\beta^0. \tag{2.4}$$

In (2.4) we drop all terms whose order is higher than the first. This means that in the total-energy operator acting on $\varphi$ we must leave only the unperturbed part symmetrized by an arbitrary Young tableau g and drop the term $\varepsilon\varphi$. Then

$$\left(H_\gamma^0 - E_i^0\right)\varphi = \sum_\beta \left(\varepsilon - V_\beta\right) C_\beta^0 \Psi_\beta^0. \tag{2.5}$$

The solution of the homogeneous analog of Eq. (2.5) for $\varphi$ is $\varphi = \Psi_\gamma^0$. But then, according to the Fredholm alternative [21], the nonhomogeneous problem (2.5) has a solution only if the vector $\langle \Psi_\gamma^0 |$ is orthogonal to the entire right-hand side:

$$\sum_\beta \left(\varepsilon \langle \Psi_\gamma^0 | \Psi_\beta^0 \rangle - \langle \Psi_\gamma^0 | V_\beta | \Psi_\beta^0 \rangle\right) C_\beta^0 = 0. \tag{2.6}$$

Thus, we have a system of equations for determining the coefficients $C_\beta^0$ of the regular zeroth wave function. The system has a solution only if

$$\left| \varepsilon \Delta_{\gamma\beta} - \langle \Psi_\gamma^0 | V_\beta | \Psi_\beta^0 \rangle \right| = 0. \tag{2.7}$$

This is the secular equation for determining the corrections to the energy. If

$$\Delta_{\gamma\beta} = \langle \Psi_\gamma^0 | \Psi_\beta^0 \rangle = \frac{1}{f_0^\gamma} \sum_{p=0}^{P} (-1)^{g_p} \langle \Phi^{P_\gamma} | \Psi_\beta^0 \rangle = \sum_{p=0}^{P} (-1)^{g_{p\gamma}} (-1)^{g_{p\beta}} \langle \Phi^0 | \Psi_\beta^0 \rangle = \frac{1}{f_0^\gamma} \sum_{p=0}^{P} (-1)^{g_{p\gamma}+g_{p\beta}},$$

$$\langle \Psi_\gamma^0 | V_\beta | \Psi_\beta^0 \rangle = \frac{1}{f_0^\gamma} \sum_{p=0}^{P} (-1)^{g_{p\gamma}} \langle \Phi^{P_\gamma} | V_\beta | \Psi_\beta^0 \rangle = \frac{1}{f_0^\gamma} \sum_{p_\gamma=0}^{P} (-1)^{g_{p\gamma}+g_{p\beta}} \langle \Phi^0 | V_\beta | \Psi_\beta^0 \rangle = \Delta_{\gamma\beta} \langle \Phi^0 | V_\beta | \Psi_\beta^0 \rangle,$$

(2.8)

the secular equation becomes

$$\prod_{\beta=1} \left( \varepsilon - \langle \Phi^0 | V_\beta | \Psi_\beta^0 \rangle \right) |\Delta_{\gamma\beta}| = 0. \tag{2.9}$$

The corrections $\varepsilon$ to the energy have definite values, $\varepsilon = \langle \Phi^0 | \hat{V}_\alpha | \Psi_\alpha^0 \rangle$, only if

$$|\Delta_{\alpha\beta}| \neq 0. \tag{2.10}$$

Then the set of zeroth wave functions antisymmetrized by Young tableaux is regular.

### 3. The Heisenberg parameter for the high Tc-materials La$_{2-x}$Me$_x$CuO$_4$ and Ba$_2$Cu$_3$O$_{6+x}$

The crystalline structure of the compound La$_2$CuO$_4$ in the stoichiometric state is depicted in Fig. 1. The crystal of copper dioxide alloyed with, say, strontium (La$_{2-x}$ Sr$_x$CuO$_4$) has a similar lattice: a body-centered tetragonal structure whose space group is 14/*mmm*. NMR and muon-precession experiments [22–24] have shown that the antiferromagnetic state occurs in this material due to the interaction of Cu$^{2+}$ ions lying in a single plane, while the interplanar magnetic interaction is weak. The magnetic form factor of the Cu$^{2+}$ ion measured in the antiferromagnetic state [25] corresponds to the $3d^9$ state. The O$^{2-}$ ion occupying a position between interacting copper ions does not affect this interaction because the electronic shell is filled.

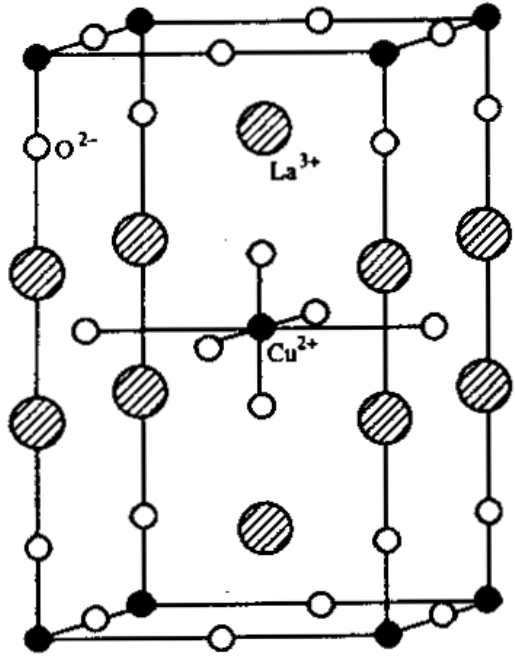

Fig.1 Crystalline and magnetic structures La$_2$CuO$_4$.

The wave function of the electrons of a pair of interacting Cu$^{2+}$ ions in the zeroth approximation corresponds to states with total spin $S=1$ or $S=0$, i.e., its spatial part is antisymmetric or symmetric, respectively. Then, from the secular equation (2.9), for the singlet and triplet states we have

$$\varepsilon_{singl,tr} = \frac{K \pm A}{1 \pm I^2},$$

where

$$\Psi_{s,a}(r_1, r_2) = \frac{1}{1 \pm I^2}\left(\psi_I(r_1)\psi_{II}(r_2) \pm \psi_I(r_2)\psi_{II}(r_1)\right),$$

$$\Delta_{ss} = 2,\ \Delta_{aa} = 2,\ \Delta_{sa} = \Delta_{as} = 0 \Rightarrow \begin{vmatrix} 2 & 0 \\ 0 & 2 \end{vmatrix} \neq 0. \qquad (3.1)$$

$$K = \left(\psi_I(r_1)\psi_{II}(r_2)\middle|V_1\middle|\psi_I(r_1)\psi_{II}(r_2)\right),$$
$$A = \left(\psi_I(r_1)\psi_{II}(r_2)\middle|V_2\middle|\psi_I(r_2)\psi_{II}(r_1)\right),$$
$$V_1 = \frac{z^2 e^2}{|R_I - R_{II}|} - \frac{ze^2}{|r_1 - R_{II}|} - \frac{ze^2}{|r_2 - R_I|} + \frac{e^2}{|r_1 - r_2|},$$
$$V_2 = \frac{z^2 e^2}{|R_I - R_{II}|} - \frac{ze^2}{|r_2 - R_I|} - \frac{ze^2}{|r_1 - R_{II}|} + \frac{e^2}{|r_1 - r_2|},$$

where $R_{I,II}$ are the radii of the interacting ions, $z$ stands for the ion charges, $r_{1,2}$ are the radius vectors of the electrons, and $I$ is the one-electron exchange density.

The wave function of an electron belonging to the Cu$^{2+}$ ion corresponds to the $3d^9_{x^2-y^2}$ orbital and is chosen in the hydrogenlike form

$$\psi(r) = \sqrt{\frac{15}{16\pi} \frac{2^7}{6!}} r^2 e^{-r} \sin^2 \vartheta \cos 2\varphi \tag{3.2}$$

(in Bohr units). Then the Heisenberg parameter for the lattice constant $R = 7.3346 a_B$ is

$$j = \varepsilon_{sing} - \varepsilon_{tr} = -0.1043278 \text{ eV}. \tag{3.3}$$

In the given case, $j$ is negative, which corresponds to antiparallel orientation of the spins at the neighboring lattice sites as being an energetically preferable configuration.

Similar calculations of j that use Eqs. (3.1) and (3.2) for the $Cu^{2+}$ ions in adjacent layers ($R = 12.3818 a_B$) yield the value

$$j = -9.673 \cdot 10^{-5} \text{ eV}, \tag{3.4}$$

which also corresponds to antiparallel orientation of the spins, but the coupling constant is very small.

Thus, the assumption that the antiferromagnetic interaction between the planes is small, which was made in the spin models of Refs. [12] and [13], is justified, and we indeed are dealing with a 2D antiferromagnetic system.

The same experiments [22–24] show that alloying with, say, strontium ($La_{2-x}Sr_xCuO_4$), causes a rapid decrease in the Néel temperature in proportion to the alloying degree $x$. The point is that alloying with a doubly charged ion of a metal activates the oxygen ion $O^{2-}$ positioned between copper ions, which becomes a single charged ion $O^{1-}$ with an unpaired electron. Now, in studying the interaction of two copper ions, we must allow for the presence of an additional electron belonging to the $O^{1-}$ ion, whose state, being highly delocalized, strongly overlaps with the electronic states of the $Cu^{2+}$ ions.

The complete electron wave function of the ion chain $Cu^{2+}- O^{1-}-Cu^{2+}$ (for the sake of brevity we denote this chain by I–II–III) can be antisymmetrized in three different ways corresponding to the following spin configurations:

$$(I\uparrow II\uparrow III\uparrow), (I\uparrow II\uparrow III\downarrow), (I\uparrow II\downarrow III\uparrow).$$

The corresponding Young tableaux are given below:

|   | Spin Young tableau | Coordinate Young tableau |
|---|---|---|
| α) | `1 2 3` | `1` / `2` / `3` |
| β) | `1 2` / `3` | `1 3` / `2` |
| γ) | `1 3` / `2` | `1 2` / `3` |

At the beginning, the electrons with numbers 1, 2, and 3 belong to the ions I, II, and III, respectively. In this case the coordinate parts of the wave functions corresponding to the Young tableaux $(\alpha), (\beta)$, and $(\gamma)$ are

$$\Psi_\alpha^0(r_1,r_2,r_3) = (1-2I_1^2+2I_2I_1^2-I_2^2)^{-1} \begin{vmatrix} \psi_I(r_1) & \psi_I(r_2) & \psi_I(r_3) \\ \psi_{II}(r_1) & \psi_{II}(r_2) & \psi_{II}(r_3) \\ \psi_{III}(r_1) & \psi_{III}(r_2) & \psi_{III}(r_3) \end{vmatrix}$$

$$\Psi_\beta^0(r_1,r_2,r_3) = (1-2I_2^2+I_1^2-I_2I_1^2)^{-1} \times$$
$$\times \begin{vmatrix} (\psi_I(r_1)\psi_{III}(r_3)+\psi_{III}(r_1)\psi_I(r_3)) & (\psi_I(r_1)\psi_{II}(r_2)+\psi_{II}(r_2)\psi_I(r_1)) \\ \psi_{III}(r_3) & \psi_{II}(r_2) \end{vmatrix}, \quad (3.6)$$

$$\Psi_\gamma^0(r_1,r_2,r_3) = (1-I_2I_1^2)^{-1} \times$$
$$\times \begin{vmatrix} (\psi_I(r_1)\psi_{II}(r_2)+\psi_I(r_2)\psi_{II}(r_1)) & (\psi_I(r_1)\psi_{III}(r_3)+\psi_I(r_3)\psi_{III}(r_1)) \\ \psi_{II}(r_2) & \psi_{III}(r_3) \end{vmatrix},$$

where we have allowed for the normalization condition (1.4) and have introduced the following notation for the integrals corresponding to the exchange densities:

$$I_1 = \int \psi_I^*(r)\psi_{II}(r)dr = \int \psi_{II}^*(r)\psi_{III}(r)dr$$
$$I_2 = \int \psi_I^*(r)\psi_{III}(r)dr. \quad (3.7)$$

For the initial distribution of the numbered electrons over the centers, the unsymmetrized perturbation operator is

$$V = \frac{z_1 z_2 e^2}{|R_I - R_{II}|} + \frac{z_1^2 e^2}{|R_I - R_{III}|} + \frac{z_1 z_2 e^2}{|R_{II} - R_{III}|} - \frac{z_2 e^2}{|r_1 - R_{II}|} - \frac{z_1 e^2}{|r_1 - R_{III}|} - \frac{z_1 e^2}{|r_2 - R_I|} - \\ - \frac{z_1 e^2}{|r_2 - R_{III}|} - \frac{z_1 e^2}{|r_3 - R_I|} - \frac{z_2 e^2}{|r_3 - R_{II}|} + \frac{e^2}{|r_1 - r_2|} + \frac{e^2}{|r_2 - r_3|} + \frac{e^2}{|r_1 - r_3|}. \quad (3.8)$$

Direct calculation of the parameter in (2.8) yields the following values:

$$\Delta_{\alpha\beta} = \Delta_{\varepsilon\gamma} = 0, \qquad \Delta_{\beta\gamma} = \frac{2}{1 - I_2 I_1^2}, \qquad \Delta_{\gamma\gamma} = \frac{4}{1 - I_2 I_1^2}, \qquad \Delta_{\alpha\alpha} = \frac{6}{1 - 2I_1^2(1 - I_2) - I_2^2},$$

$$\Delta_{\beta\beta} = \frac{4}{1 - I_1^2 + I_2^2 - I_2 I_1^2}.$$

The determinant condition (2.10) has the form

$$\begin{vmatrix} \Delta_{\alpha\alpha} & 0 & 0 \\ 0 & \Delta_{\beta\beta} & \Delta_{\beta\gamma} \\ 0 & \Delta_{\gamma\beta} & \Delta_{\gamma\gamma} \end{vmatrix} \neq 0. \quad (3.9)$$

Thus, condition (2.10) is met. In this case and in accordance with (2.9), the corrections to the energy are

$$\varepsilon_1 = \left(\Phi^0(r_1 r_2 r_3)|V_\alpha|\Psi_\alpha^0\right),$$
$$\varepsilon_2 = \left(\Phi^0(r_1 r_2 r_3)|V_\beta|\Psi_\beta^0\right), \quad (3.10)$$
$$\varepsilon_3 = \left(\Phi^0(r_1 r_2 r_3)|V_\gamma|\Psi_\gamma^0\right),$$

where

$$\left|\Phi^0(r_1 r_2 r_3)\right) = \left|\psi_I(r_1)\psi_{II}(r_2)\psi_{III}(r_3)\right).$$

Note that in contrast to the two-center case, superexchange three-center integrals of the form

$$K_{I \to II, II \to III} = \iint \frac{\psi_I^*(\vec{r}_1)\psi_{II}^*(\vec{r}_2)\psi_{II}(\vec{r}_1)\psi_{III}(\vec{r}_2)}{|\vec{r}_1 - \vec{r}_2|} d^3 r_1 d^3 r_2 \quad (3.11)$$

contribute substantially to all the expressions in (3.10). Such integrals ensure three-center correlation of the spins, since they can enter into in the general expressions (3.10) for the energy with different signs. The sign sequence of these superexchange terms is determined by the Young diagrams used in antisymmetrizing the wave function in (3.10). Due to this superexchange interaction, long-range order may set in the system without the conduction electrons participating, as is the case in the Kondo and RKKY models.

Calculations of the matrix elements in (3.10) for the given lattice constant $R = 3.88$ Å with allowance for the superexchange contributions (see Appendix B) yield

$$\varepsilon_1(\uparrow\uparrow\uparrow) = -0.763 \text{ a.u.}$$
$$\varepsilon_2(\uparrow\downarrow\downarrow) = -0.671 \text{ a.u.} \qquad (3.12)$$
$$\varepsilon_3(\uparrow\downarrow\uparrow) = -0.639 \text{ a.u.}$$

In the given case of $Cu^{2+}$ ions, the orientation of the spins is ferromagnetic, with the Heisenberg parameter being

$$J \approx \varepsilon_2 - \varepsilon_1 = 0.092 \text{ a.u.} = 2.513 \text{ eV}. \qquad (3.13)$$

Comparison with (3.3) yields the following value of the parameter ratio:

$$\frac{J}{|j|} = 25. \qquad (3.14)$$

Thus, alloying the material, which activates the oxygen ions $O^{1-}$, does indeed reorient the electron spins in $Cu^{2+}$ and leads to strong ferromagnetism. Hence, as Birgeneau and Schirane [25] pointed out by analyzing the experimental facts, a strong ferromagnetic bond in the $CuO_2$ plane destroys the local antiferromagnetic order. In the case of strong localization, the concentration of $\Phi$-bonds would be equal to $x$. As $x$ grows, the localization length $l_0$ of each hole increases, which leads to an increase in the effective concentration of $\Phi$-bonds. Birgeneau and Schirane [25] found that a large value of $J/|j|$ reduces the threshold values of $x$ at which antiferromagnetism in alloyed $La_2CuO_4$ disappears even if $x$ is small [26].

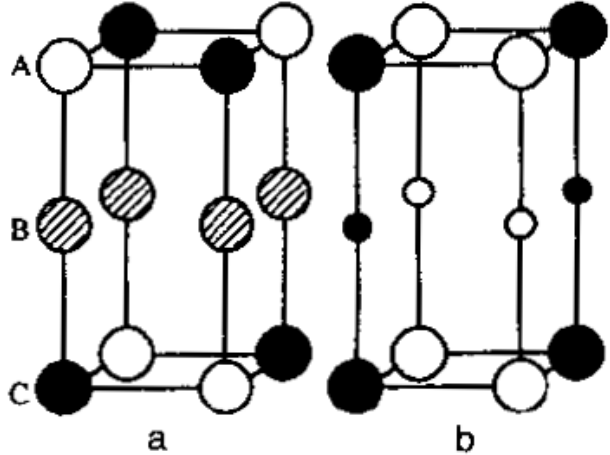

FIG. 2. (a) Magnetic spin structure of $YBa_2Cu_3O_{6+x}$ with $x=0$. Only copper atoms are depicted. The hatched circles stand for the nonmagnetic $Cu^{1+}$ ions, the dark and light circles stand for the antiparallel spins in the $Cu^{2+}$ positions, and the solid straight lines denote double bonds with oxygen atoms. (b) The second type of spin structure observed for large values of $x$. The average spin in layer B is the fraction $\varepsilon$ of the spin in a $CuO_2$ layer.

The experimental data on another copper dioxide, YBa$_2$Cu$_3$O$_{6+x}$ suggests a remarkable resemblance between the two systems [4, 27]. The sublattice in YBa$_2$Cu$_3$O$_6$ is depicted in Fig. 2. Each chemical cell contains two CuO$_2$ layers, denoted by A and C. The copper atoms in layer B have no bonds with oxygen. X-ray absorption measurements have clearly shown that the valence of the copper atoms in layer B is 1+, so that the atoms are nonmagnetic.

To represent the spin structure of the Cu$^{2+}$ ions in layers A and C, the antiparallel spins are depicted by dark and light circles. The calculation of the corresponding parameter $j$ is done by (3.1) – (3.3) with a lattice constant $R = 7.225 a_B = 3.822$ Å:

$$j_{AA} = \varepsilon_{sing} - \varepsilon_{tr} = -0.0935 \, \text{eV}$$

inside a layer, and

$$j_{AC} = \varepsilon_{sing} - \varepsilon_{tr} = -0.0713 \, \text{eV}$$

between layers A and C. An estimate of the same parameter done by analyzing the experimental data on the Néel temperature yielded $j \approx 0.086$ eV. X-ray absorption measurements have shown that adding oxygen facilitates the transition of Cu$^{1+}$ into Cu$^{2+}$. Now the Cu$^{2+}$ in layer B facilitates the destruction of the antiferromagnetic order between neighboring Cu$^{2+}$ ions in layers A and C.

Consider a system of three Cu$^{2+}$ ions, with the layers A, B, and C each having one ion. Then, using Eqs. (3.5) – (3.10), we can calculate the energy corrections to the spin configuration in the Young diagram:

$$\varepsilon_1(\uparrow\uparrow\uparrow) = -0.33847 \text{ a.u,}$$
$$\varepsilon_2(\uparrow\downarrow\downarrow) = -0.33027 \text{ a.u,}$$
$$\varepsilon_3(\uparrow\downarrow\uparrow) = -0.34666 \text{ a.u.}$$

Thus, the states with ferromagnetic orientation of the spins in layers A and B are the most probable, the spin in layer B may be assumed fluctuating, and the complete ferromagnetic state has a Heisenberg parameter

$$J_1 = \varepsilon_2 - \varepsilon_1 = 8.2 \cdot 10^{-3} E_B = 0.223 \, \text{eV}$$

while the state of the oppositely directed spin in layer B has a Heisenberg parameter

$$J_2 = 16.4 \cdot 10^{-3} E_B = 0.446. \, \text{eV}$$

The average value of the magnetic moment per magnetic atom is

$$m = \frac{3\mu_0 + \mu}{2 \cdot 3} = \frac{2}{3}\mu_0 = 0.6(6)\mu_0.$$

The experiment of Birgeneau *et al.* [28] gives the same value for the maximally ordered moment.

## Conclusion

We have developed a variant of exchange perturbation theory that allows for degeneracy in total spin, with the applicability criteria (1.18) and (2.10) added. Using it we have derived a procedure for *ab initio* calculations of the Heisenberg parameter for high-*Tc* materials, a procedure that is based on first principles and avoids computer simulations, so that the necessary relationships are obtained in analytical form. Numerical estimates of the energy values for given lattice parameters yield results that are in good agreement with those of experimental and phenomenological approaches. For instance, for La$_2$CuO$_4$ the experiment of Peters *et al.* [29] yields the following values of the Heisenberg parameter.

For the antiferromagnetic interaction of the ions in Cu$^{2+}$–Cu$^{2+}$ lying in a single plane, $|J_1|= 0.116\,\text{eV}$; between the planes, $|J_2|= 2\cdot 10^{-6}\,\text{eV}$.

For the ferromagnetic Cu$^{2+}$–O$^-$–Cu$^{2+}$, interaction $J \approx 1\,\text{eV}$; our values of the corresponding parameters are $|J_1|= 0.104\,\text{eV}$, $|J_2|\approx 9\cdot 10^{-5}\,\text{eV}$ $|J|\approx 2.5\,\text{eV}$.

For YBa$_2$Cu$_3$O$_6$, the Heisenberg energy estimated using an analysis of the Néel temperature is $|J|\sim 0.086\,\text{eV}$ (Refs. [24] and [27]); our values are $|J|\approx 0.0935\,\text{eV}$ inside a layer and $|J|\approx 0.07\,\text{eV}$ between layers.

Note that the following simplifications were made in specific calculations:
1) Intratomic interaction is incorporated only indirectly by using one-electron states, whose parameters are taken from the Gomba´s and Szondy tables [30].
2) The effect of the mean crystalline field on the intercenter interaction of *d*-electrons is ignored; the latter, however, incorporates the interaction with the nearest-neighbor ions.
3) The influence of conduction electrons on the magnetic effects due to the orientation of spins of *d*-electrons is ignored.
4) Finally, spin-orbit coupling effects is ignored.

In conclusion we note that the analytical expressions for the energy of a three-center chain of atoms contains ~as the leading contribution! the nonadditive contribution of the three-center interaction (see Appendix B). The terms are $K_{1321}$, $K_{1232}$, $K_{2321}$, $K_{1323}$, $K_{2313}$, $K_{1231}$, $K_{2312}$, and $K_{1332}$. Such integrals determine the structure, since they depend not only on the intercenter distance but also on the angles between the straight lines connecting these centers. We believe that they are responsible for exchange-correlation effects in solids, including spin systems.

We are grateful to V. V. Rumyantsev for his attention to our work and to his critical remarks, which enabled us to significantly improve the presentation in this paper. The work was made possible by the financial support of the Russian Academy of Sciences (Young Scientist Stipend of the Russian Academy of Sciences) given to one of the coauthors.

# Appendix A: Completeness property of the nonorthogonal basis of antisymmetric functions

States that are antisymmetric under particle permutations are nonorthogonal. Nevertheless, they constitute complete system. To verify this, we act with the operator $\sum_n |\Phi_n\rangle\langle\Psi_n|$ on an arbitrary function antisymmetric according to the same Young diagram as the states $|\Psi_n\rangle$:

$$\sum_n |\Phi_n\rangle\langle\Psi_n|\Psi\rangle = \sum_n \sum_{p=0}^{P} |\Phi_n\rangle\langle\Phi_n^p|\Psi\rangle \cdot (-1)^{g_p} \frac{1}{f_n} \tag{A1}$$

where

$$f_n = \sum_{p=0}^{P} \langle \Phi_n | \Phi_n^p \rangle (-1)^{g_p}$$

Using the antisymmetry of the vector $|\Psi\rangle$, we can write

$$\sum_n \sum_{p=0}^{P} \frac{1}{f_n} |\Phi_n\rangle \langle \Phi_n^p | \Psi \rangle = \sum_n \sum_{p=0}^{P} \frac{1}{f_n} |\Phi_n\rangle \langle \Phi_n | \Psi \rangle (-1)^{g_p} = $$
$$= \sum_{p=0}^{P} \frac{1}{f_0} |\Psi\rangle = \frac{P}{f_0} |\Psi\rangle, \qquad (A2)$$

where we have used the fact that $f_n \approx f_0 = \sum_{p=0}^{P} \langle \Phi_0 | \Phi_0^p \rangle (-1)^{g_p}$ and the completeness property of an orthogonal basis of nonsymmetric functions, $\sum_n |\Phi_n\rangle \langle \Phi_n| = 1$

Thus, we have

$$\frac{f_0}{P} \sum_n |\Phi_n\rangle \langle \Psi_n | = \sum_n |\Phi_n\rangle \langle \Phi_n |, \qquad (A3)$$

which is simply the completeness property of the system. Using (A3), we can decompose an arbitrary antisymmetrized state in the same antisymmetric states.

In deriving the completeness property we assumed that $f_0/f_n \approx 1$. Let us now estimate the smallness of the terms discarded. Clearly, the above ratio can be written as

$$\frac{f_0}{f_n} = 1 + \sum_{p=0}^{P} (-1)^{g_p} \{ \langle \Phi_0 | \Phi_0^p \rangle - \langle \Phi_n | \Phi_n^p \rangle \} + O(I^{2P}), \qquad (A4)$$

where

$$\mathrm{Supr}\left( \sum_{p=0}^{P} (-1)^{g_p} \left[ \langle \Phi_0 | \Phi_0^p \rangle - \langle \Phi_n | \Phi_n^p \rangle \right] \right) \approx \frac{N!}{g_P!(N-g_P)!} \cdot \frac{I_0 - I_n}{1 + I_0 + I_n}$$

$I_0$ and $I_n$ are still the one-electron exchange densities of the ground and excited states, and $N$ and $g_P$ are the respective total number of electrons and the number of electrons participating in the intercenter permutation.

Since the overlap integrals are, in general, the product of a polynomial, $P(R/a_B)$, and an exponential, $\exp(-R/a_B)$, the discarded terms are

$$\mathrm{Supr}\left\{ \frac{N!}{g_P!(N-g_P)!} P\left(\frac{R}{a_B}\right) \exp\left(-\frac{R}{a_B}\right) \right\} = \mathrm{const} < 1 \qquad (A5)$$

This constant can, in principle, be accounted for in (A3).

# Appendix B: Expression for obtaining the corrections to the energy

1. The energies of a two-center interaction are

$$E_{\text{sing}} = \frac{e^2}{1+I^2}\left[\frac{z_1^2}{R}(1+I^2) + 2z_1 C_{11} - 2z_2 C_{12} + K_{1212} + K_{1221}\right], \quad \text{(B1)}$$

$$E_{\text{tr}} = \frac{e^2}{1-I^2}\left[\frac{z_1^2}{R}(1-I^2) - 2z_1 C_{11} + 2z_2 C_{12} + K_{1212} - K_{1221}\right], \quad \text{(B2)}$$

where we have introduced the following notation:

$$C_{ij} = \int \psi^*_i(\vec{r})\psi_j(\vec{r})\frac{d^3r}{|\vec{r}-\vec{R}|}$$

is the direct or exchange interaction of an electron with a ''foreign'' nucleus, $i$ and $j$ label the nucleus,

$$K_{ijkl} = \int \frac{\psi^*_i(\vec{r}_1)\psi^*_k(\vec{r}_2)\psi_j(\vec{r}_1)\psi_l(\vec{r}_2)}{|\vec{r}_1-\vec{r}_2|}d^3r_1 d^3r_2$$

is the exchange interaction of two electrons distributed between the nuclei $i, j, k, l$, and

$$K_{ijij} = \int \frac{\psi^*_i(\vec{r}_1)\psi^*_j(\vec{r}_2)\psi_i(\vec{r}_1)\psi_j(\vec{r}_2)}{|\vec{r}_1-\vec{r}_2|}d^3r_1 d^3r_2$$

is the direct Coulomb interaction of two electrons centered at different nuclei $i$ and $j$.

2. The energies of the three-center interaction are

$$E_1 = \frac{1}{6}\Delta_{\alpha\alpha}(K_0 - K_{2-3} - K_{1-2} + K_{1-2,1-3} + K_{1-2,2-3} - K_{1-3}), \quad \text{(B3)}$$

$$E_2 = \frac{1}{4}\Delta_{\beta\beta}(K_0 + K_{1-3} - K_{2-3} - K_{1-2,2-3}), \quad \text{(B4)}$$

$$E_3 = \frac{1}{4}\Delta_{\gamma\gamma}(K_0 - K_{2-3} + K_{1-2} - K_{1-2,1-3}), \quad \text{(B5)}$$

where

$$K_0 = \frac{z_1 z_2}{R} + \frac{z_1^2}{R} - z_1(C_{11} - C_{22} + S_{22} + S_{33}) - z_2(B_{11} - B_{33}), \quad \text{(B6)}$$

$$K_{1-2} = \left(\frac{z_1 z_2}{R} + \frac{z_1^2}{R}\right)I_1^2 - (z_2 I_1 B_{21} + z_1 I_1 C_{21} + z_1 I_1 S_{12} + z_1 I_1 C_{12} + \\ z_1|I_1|^2 S_{33} + I_1^2 z_2 B_{33})z_1 + K_{1221} + K_{2313} I_1 + K_{1323} I_1. \quad \text{(B7)}$$

and $K_{2-3}$ can be obtained from (B7) by cyclic permutation of the subscripts 1, 2, and 3.

Similarly, for $K_{1-3}$ we have

$$K_{1-2,2-3} = \left(\frac{z_1 z_2}{R} + \frac{z_1^2}{R}\right) I_1^2 I_2 - (z_2 I_1 I_2 B_{23} + z_1 I_1 I_2 C_{32} + z_1 I_1^2 S_{13} + z_1 I_1^2 C_{13} + $$
$$+ z_1 I_1 I_2 S_{21} + I_1 I_2 z_2 B_{21}) z_1 + K_{1231} I_1 + K_{2312} I_2 + K_{1332} I_1. \quad \text{(B8)}$$

Here we have used the integrals

$$B_{ij} = \int \frac{\psi_i^*(\vec{r}) \psi_j(\vec{r})}{|\vec{r} - \vec{R}|} d^3 r,$$

$$S_{ij} = \int \frac{\psi_i^*(\vec{r}) \psi_j(\vec{r})}{|\vec{r}|} d^3 r$$

complementary to (B1) and (B2).